\documentclass[prd,superscriptaddress,preprint]{revtex4}

\usepackage{mathrsfs,amsmath}
\usepackage{eurosym}
\usepackage{amssymb}
\usepackage{bbold}

\usepackage{latexsym,epsfig,epstopdf}
\usepackage{amssymb,amsmath,amsthm}

\usepackage{times}
\usepackage{color}

\usepackage{lmodern}
\usepackage[utf8]{inputenc}

\usepackage{graphicx}

\usepackage{color}

\usepackage{braket}

\usepackage[mathscr]{euscript} 

\def\be{\begin{equation}}
\def\ee{\end{equation}}
\def\bea{\begin{eqnarray}}
\def\eea{\end{eqnarray}}

\begin{document}




%

\title{Connections between Weyl geometry, quantum potential and quantum entanglement
}

\author{Shi-Dong Liang}
\email{stslsd@mail.sysu.edu.cn}
\affiliation{School of Physics, Sun Yat-Sen University, Guangzhou 510275, China}
\affiliation{State Key Laboratory of Optoelectronic Material and Technology, and\\
Guangdong Province Key Laboratory of Display Material and Technology,
Sun Yat-Sen University, Guangzhou, 510275, China}

\author{Wenjing Huang}
\affiliation{School of Physics, Sun Yat-Sen University, Guangzhou 510275, China}

\begin{abstract}
The Weyl geometry promises potential applications in gravity and quantum mechanics. We study the relationships between the Weyl geometry, quantum entropy and quantum entanglement based on the Weyl geometry endowing the Euclidean metric. We give the formulation of the Weyl Ricci curvature and Weyl scalar curvature in the $n$-dimensional system. The Weyl scalar field plays a bridge role to connect the Weyl scalar curvature, quantum potential and quantum entanglement. We also give the Einstein-Weyl tensor
and the generalized field equation in 3D vacuum case, which reveals the relationship between Weyl geometry and quantum potential.
Particularly, we find that the correspondence between the Weyl scalar curvature and quantum potential is dimension-dependent and works only for the 3D space, which reveals a clue to quantize gravity and a understanding why our space must be 3D if quantum gravity is compatible with quantum mechanics.
We analyze numerically a typical example of two orthogonal oscillators to reveal the relationships between the Weyl scalar curvature, quantum potential and quantum entanglement based on this formulation. We find that the Weyl scalar curvature shows a negative dip peak for separate state but becomes a positive peak for the entangled state near original point region, which can be regarded as a geometric signal to detect quantum entanglement.

\end{abstract}


\maketitle


\section{Introduction}
Since Einstein's general relativity has been successful in gravity theory, one has realized physics behind mathematics beyond Euclidean geometry.\cite{Charles}
The gravity can be understood by spacetime curvature in pseudo-Riemannian geometry and the Universe evolution was started from a the big bang.\cite{Charles} However,
the recent observations of high redshift supernovae and the Boomerang/Maxima/WMAP data showed that the content of the Universes resides in two unknown dark matter and dark energy,\cite{Supernova} which inspires a lot of attempts to understand what new physics behind of dark matter and dark energy and what gravity theory beyond the Einstein's general relativity. \cite{Wu,Xu,Haghani} There have been many efforts of generalized Einstein's gravity such as so-called Palatini formulation, Weyl-Cartan theory and Weyl-Cartan-Weitzenb$\ddot{o}$ck  theory based on Weyl-Cartan geometry for understanding dark matter and dark energy.\cite{Wu,Xu,Haghani}
The spacetime geometry and gravity can be understood by more deep geometric variables, such as torsion and non-metricity. \cite{Wu,Xu,Haghani}
These generalized gravity theory imply some deep connections between physics and geometry beyond Riemann geometry. Especially, the non-metricity of Weyl geometry associates a scalar field curving space, which could provide a new insight to novel phenomena of cosmology and connecting quantum world.\cite{Santamato,Carroll}

On the other hand, the Yang-Mills gauge theory can be traced back Weyl geometry.\cite{Milutin,Borchers} The geometric approach has been also extended to apply in condensed matter physics recently.\cite{Kane,Hasan,Qi,Chiu} The quantum states such as quantum Hall effect and topological insulator states can be characterized by the geometric approaches, such as winding number, Berry curvature and Chern number, which are associated with topological invariant.\cite{Kane,Hasan,Qi,Chiu}

Recently one found some connections between quantum potential and Weyl scalar curvature based on the hydrodynamical representation of quantum mechanics.\cite{Carroll,Santamato,Novello,Licata1} The Weyl geometry endows the covariant derivative of the metric with a scalar field that describes the vector length varying when the vector moves in Weyl manifold.
In particular, when a special Weyl geometry is considered to endow the $3+1$-Weyl manifold with the three-dimensional (3D) spatial Euclidean metric, which is so-called Q-Wis space,\cite{Novello,Licata1} the quantum potential plays an analog with the Weyl scalar curvature for the three-dimensional (3D) system,\cite{Novello,Licata2} which also connect to quantum entropy and Fisher information. \cite{Licata2,Fiscaletti1,Fiscaletti2,Licata3,Castro} The hydrodynamical representation of Sch\"{r}odinger equation can be derived from the geometric variational principle in the Q-Wis geometry. These results give us a hint to explore the connection between quantum entanglement and Weyl geometry.
Especially is the Weyl scalar curvature endowed with quantum potential working in the 3D space coincidental or a hint for some clue for understanding deeper fundamental puzzles? A natural question is this connection between the Weyl scalar curvature and quantum potential working beyond the 3D space, which could inspire us to understand why our space must be 3D and how can we quantize gravity.

In this paper, we will generalize the formulation of the 3D Q-Wis Weyl geometry for quantum mechanics to that of the $n$-dimensional (nD) case, including the Weyl Ricci curvature, Weyl scalar curvature and Einstein-Weyl tensor which are expressed in terms of the Weyl scalar field in Sec. II. Then we will derive the hydrodynamical representation of Sch\"{r}odinger equation by the geometric variational principle in Sec. III. In Sec. VI, we will give the relationships between the Weyl scalar curvature, quantum entropy,  quantum potential, and quantum entanglement as well as the generalized Einstein-Weyl field equation. We will discuss the dimensional dependent connections between the Weyl scalar curvature and quantum potential, which will give us a novel insight to understand our spatial dimension and how to quantize gravity.
As an typical example we will analyze the relationship between the Weyl scalar curvature and quantum entanglement for two orthogonal linear oscillators in Sec. V.
We will find that the Weyl scalar curvature can play a role of indicator or signal to detect the quantum entanglement of quantum states.
Finally we will give conclusions and outlook.

\section{Weyl geometry}
\subsection{Brief review on Weyl geometry}
The differential geometric manifold is characterized by the triplet elements, $(g,\nabla,\mathcal{M})$, where $g$ is metric, $\nabla$ is connection and $\mathcal{M}$ is non-metricity. The geometric properties of a manifold are completely determined by the transplantation law of vectors. The arc length joining two adjacent points $x^i$ and $x^{i}+dx^{i}$ is defined $ds^2=g_{ik}dx^{i}dx^{k}$ with positive-definite metric tensor $g_{ik}$. The variation of $\delta\xi^i$ of a vector $\xi^i$ in a transplantation from $x^i$ to $x^{i}+dx^{i}$ is given by $\delta\xi^i=\Gamma^{i}_{k\ell}\xi^{k}dx^\ell$, where $\Gamma^{i}_{k\ell}$ is the affine connections on manifold.

In Riemannian geometry, the transplantation of vectors holds the length of vectors, namely $\nabla_{\gamma} g_{\mu\nu}=0$.
Weyl geometry allow the length of vectors varying in their transplantation. The length of a vector is defined by
$\ell=\left(g_{ik}\xi^{i}\xi^{k}\right)^{1/2}$, which varies based on the linear law, $\delta\ell=\ell\phi_{k}dx^k$, where
$\phi_{k}$ is the components of a scalar field. (called Weyl scalar field $\phi_{\beta}$).
This properties is called non-metricity. By using similar approach to Riemanian geometry, the connection coefficient in Wely geometry without torsion is obtained\cite{Santamato}
\begin{equation}\label{WC1}
\Gamma^{\alpha}_{\beta\gamma}
=-\left\{^{\alpha}_{\beta\gamma}\right\}+g^{\sigma\alpha}\left(g_{\sigma\beta}\phi_{\gamma}
+g_{\sigma\gamma}\phi_{\beta}-g_{\beta\gamma}\phi_{\sigma}\right),
\end{equation}
where $\left\{^{\alpha}_{\beta\gamma}\right\}$ denotes the Christoffel symbols defined by the metric tensor $g_{\alpha\beta}$ and its derivatives. The covariant derivative of vectors $\xi^{k}$ is defined by
\begin{equation}\label{CD1}
\nabla_{\ell} \xi^{k}=\partial_{\ell}\xi^{k}+\Gamma^{k}_{\ell \alpha}\xi^{\alpha},
\end{equation}
The curvature tensor is introduced by
\begin{equation}\label{RC1}
\left(\nabla_{k}\nabla_{\ell}- \nabla_{\ell}\nabla_{k}\right)\xi^{\alpha}=\mathcal{R}^{\alpha}_{mk\ell}\xi^{m},
\end{equation}
where $\mathcal{R}^{\alpha}_{mk\ell}$ is the curvature tensor in Weyl geometry, which has the same formula of Riemanian geometry,
\begin{equation}\label{RC2}
\mathcal{R}^{\alpha}_{mk\ell}=\partial_{k}\Gamma^{\alpha}_{m\ell}-\partial_{\ell}\Gamma^{\alpha}_{mk}
+\Gamma^{\alpha}_{n\ell}\Gamma^{n}_{mk}-\Gamma^{\alpha}_{nk}\Gamma^{n}_{m\ell},
\end{equation}
The curvature tensor $R^{\alpha}_{mk\ell}$ obeys the same symmetric properties as the curvature tensor of Riemannian geometry. Similarly, the
Ricci symmetric tensor is defined by $\mathcal{R}_{m\ell}=\mathcal{R}^{\alpha}_{m\alpha\ell}$ and the scalar curvature is defined by $\mathcal{R}=g^{m\ell}\mathcal{R}_{m\ell}$. By using (\ref{RC2}) the Weyl scalar curvature can be obtained\cite{Santamato}
\begin{equation}\label{SC1}
\mathcal{R}=R+(n-1)(n-2)\phi_{k}\phi^{k}-2(n-1)\partial_{k}\phi^{k},
\end{equation}
where $R$ is the Riemannian curvature in terms of the Christoffel symbols.

\subsection{Q-Wis geometry}
In order to reveal the connection between the geometric objects and quantum effects of quantum systems, let us consider a special Weyl geometry endowed on the Euclidean metric. The length of vector varies by the linear law $\delta\ell=\ell \phi_{a}dx^{a}$, where $\phi_{a}:=\frac{\partial f}{\partial x^a}$.
It should be noted that we define the Weyl scalar field $\phi_{k}$ is the derivative of another scalar field $f$.
We will call $f$ as the Weyl scalar field hereinafter. This geometric setting is called
the Q-wis geometry.\cite{Novello,Licata1}
The covariant derivative of the metric is expressed as
\begin{equation}\label{QWM1}
\nabla_{k}g_{ab} = \frac{\partial f}{\partial x^k}g_{ab}.
\end{equation}
Since the manifold is endowed on Euclidean metric, the length $\ell$ of vectors moves along a closed path $\oint d\ell=0$.
The affine connection defined on the tangent space of the manifold $TM$,
\begin{equation}\label{AC2}
\nabla_{a}\xi_{b} = \frac{\partial \xi_{a}}{\partial x^b}+\left\{^{k}_{ab}\right\}\xi_{k},
\end{equation}
where $\xi_{a}\in TM$ and $\left\{^{k}_{ab}\right\}$ is the Christoffel symbol. It does not vanish and depends on the Weyl scalar field.\cite{Novello,Licata1}

Let us consider the n-dimensional(nD) Q-wis geomery, taking the covariant derivative of the metric in (\ref{QWM1}) into account, the Christoffel symbol can be expressed as
\begin{equation}\label{WC1}
\left\{^{k} _{ab}\right\}
= -\frac{1}{2}\left(\delta_{a}^{k}\frac{\partial f}{\partial x^b}+\delta_{b}^{k}\frac{\partial f}{\partial x^a}-g_{ab}g^{kc}\frac{\partial f}{\partial x^c}\right).
\end{equation}
Notice that $\delta^{a}_{a}=n$ for the n-dimensional systems, by the similar computation to Riemannain geometry, the Weyl-Ricci curvature is obtained
\begin{equation}\label{RiC1}
\mathcal{R}_{ij} =\frac{1}{2} \left[(2-n)\frac{\partial^{2}f}{\partial x^{i}\partial x^{j}}-g_{ij}\frac{\partial^{2}f}{\partial x^{k}\partial x_{k}}\right]
+\frac{2-n}{4}\left(\frac{\partial f}{\partial x^{i}}\frac{\partial f}{\partial x^{j}}-g_{ij}\frac{\partial f}{\partial x^{k}}\frac{\partial f}{\partial x_{k}}\right).
\end{equation}
The Weyl scalar curvature defined by $\mathcal{R}=g^{ij}\mathcal{R}_{ij}$ is obtained
\begin{equation}\label{SR1}
\mathcal{R} =\frac{1-n}{2} \left(2\frac{\partial^{2}f}{\partial x^{k}\partial x_{k}}
+\frac{2-n}{2}\frac{\partial f}{\partial x^{k}}\frac{\partial f}{\partial x_{k}}\right).
\end{equation}
Thus, the Weyl Ricci curvature and Weyl scalar curvature can be expressed in terms of the Weyl scalar field.
In order to explore the physical meaning of the Weyl scalar field, let us define the Weyl scalar field that is proportional to the Shannon-type entropy, $f:=-\lambda\ln\Omega$, where $\Omega=\sqrt{\rho}=|\psi|$.\cite{Novello,Licata1}
We identify $\rho$ to be the probability density of particles and
$\psi$ is the wave function of particles in the $L^2(\mathbb{R}^n)$ Hilbert space. The Weyl-Ricci curvature for the $n$-dimensional systems can be rewritten as
\begin{eqnarray}\label{RiC2}
\mathcal{R}_{ij} &=& \frac{\lambda}{2} \left[\left(1+\frac{\lambda}{2}\right)\frac{2-n}{\Omega^2}
\frac{\partial \Omega}{\partial x^{i}}\frac{\partial \Omega}{\partial x^{j}}-\frac{2-n}{\Omega}
\frac{\partial^{2}\Omega}{\partial x^{i}\partial x^{j}}
- \left(1+\frac{2-n}{2}\lambda\right)\frac{g_{ij}}{\Omega^2}
\frac{\partial \Omega}{\partial x^{k}}\frac{\partial \Omega}{\partial x_{k}}\right. \nonumber\\
&+&\left.\frac{g_{ij}}{\Omega}\frac{\partial^{2}\Omega}{\partial x^{k}\partial x_{k}}\right],
\end{eqnarray}
and the Weyl scalar curvature is expressed as
\begin{equation}\label{SR2}
\mathcal{R} =(1-n)\lambda \left[\left(1+\frac{2-n}{4}\lambda\right)
\frac{1}{\Omega^2}\frac{\partial \Omega}{\partial x^{k}}\frac{\partial \Omega}{\partial x_{k}}
-\frac{\partial^{2}\Omega}{\Omega\partial x^{k}\partial x_{k}}\right].
\end{equation}

Similarly to Riemanian geometry, the Einstein-Weyl tensor is defined by $\mathcal{G}_{ij}=\mathcal{R}_{ij}-\frac{1}{2}g_{ij}\mathcal{R}$ in Q-Wis geometry, where we ignore the spatial curvature for giving quantum effect. The Einstein-Weyl tensor can be obtained
\begin{eqnarray}\label{GE1}
\mathcal{G}_{ij}&=&\frac{\lambda(\lambda+2)(2-n)}{4}\frac{\partial_{i}\Omega\partial_{j}\Omega}{\Omega^2}-\frac{(2-n)\lambda}{2}
\left(\frac{\partial_{i}\partial_{j}\Omega}{\Omega}-g_{ij}\frac{\partial^{k}\partial_{k}\Omega}{\Omega}\right) \nonumber\\
&+&\frac{\lambda(n-2)(3\lambda-n\lambda+4)}{8}g_{ij}\frac{\partial^{k}\Omega\partial_{k}\Omega}{\Omega^2}.
\end{eqnarray}

The Einstein-Weyl scaler defined by $\mathcal{G}=g^{ij}\mathcal{G}_{ij}$ in Q-Wis geometry can be also obtained
\begin{equation}\label{GS1}
\mathcal{G}=\frac{\lambda(2\lambda-n\lambda+4)(n^2-3n+2)}{8}\frac{\partial^{k}\Omega\partial_{k}\Omega}{\Omega^2}
-\frac{\lambda(n-1)(n-2)}{2}\frac{\partial^{k}\partial_{k}\Omega}{\Omega}.
\end{equation}
Interestingly, the Einstein's tensor and scaler depends on the dimension of space. The Einstein's scalers from the 1D to 3D cases are reduced to
\begin{equation}\label{GS}
\mathcal{G} = \left\{
\begin{array}{cc}
 0, & \quad \textrm{for}\quad n=1,2 \\
\frac{\lambda(4-\lambda)}{4}\frac{\partial^{k}\Omega\partial_{k}\Omega}{\Omega^2}
-\lambda\frac{\partial^{k}\partial_{k}\Omega}{\Omega}, & \quad \textrm{for}\quad n=3
\end{array}.
\right.
\end{equation}
It implies that there does not exist quantum effects in the 1D and 2D space for Q-Wis geometry. In other words, quantum fluctuation does not curve the 1D and 2D space. This maybe tells us why our space must be higher than 2D dimension.

For 3D vacuum case, the Einstein-Weyl field equation in Q-Wis geometry can be given by
\begin{equation}\label{GE2}
\frac{\partial_{i}\partial_{j}\Omega}{\Omega}-\frac{\lambda+2}{2}\frac{\partial_{i}\Omega\partial_{j}\Omega}{\Omega^2}
+g_{ij}\left(\frac{\partial^{k}\Omega\partial_{k}\Omega}{\Omega^2}-\frac{\partial^{k}\partial_{k}\Omega}{\Omega}\right)=0.
\end{equation}

\subsection{Quantum hydrodynamics in Q-wis geometry}
From the dynamical point of views, the equation of motion is derived from the extreme of action by variational principle. The action of dynamics in the Q-wis geometry is constructed\cite{Novello,Licata1}
\begin{equation}\label{At1}
I=\int dtd^{3}x \sqrt{g}\Omega^{2}(\eta^{2}\mathcal{R}-\mathcal{L}_{m}),
\end{equation}
where $\eta$ is the parameter and $\mathcal{R}$ is the Weyl scalar curvature. The $\mathcal{L}_{m}$ is the Lagrangian of particles defined by
\begin{equation}\label{Lg1}
\mathcal{L}_{m}=\frac{\partial S}{\partial t}+\frac{1}{2m}\nabla S\cdot \nabla S +V,
\end{equation}
where $\nabla S:=\mathbf{v}$ is the velocity of particles in the hydrodynamical representation of quantum mechanics, namely $\psi=\sqrt{\rho}e^{iS/\hbar}$. By the variational principle, varying $\Omega$ and $S$, we can obtain the equation of motion.
\begin{subequations}\label{SchE1}
\begin{eqnarray}
\frac{\partial S}{\partial t}+\frac{1}{2m}\nabla S\cdot \nabla S +V-\eta^{2}\mathcal{R} &=& 0 \\
\frac{\partial \Omega^2}{\partial t} +\nabla \left(\Omega^{2}\frac{\nabla S}{m}\right)&=& 0.
\end{eqnarray}
\end{subequations}
One can identify the last term $\eta^{2}\mathcal{R}$ in (\ref{SchE1}a) to be quantum potential by choosing the parameter $\eta$.\cite{Novello} Thus, the equation of motion (\ref{SchE1}) is the hydrodynamical representation of Sch\"{r}odinger's equation, where
the Weyl scalar curvature plays a role of quantum potential.
Note that the Weyl scalar curvature depends on the Christoffel symbol, by varying the Christoffel symbol in the action (\ref{At1}), we can obtain the metric compatible condition for a Q-wis geometry, $\nabla_{k} g_{ab}=-4\frac{\partial \ln\Omega}{\partial x^{k}}g_{ab}$.\cite{Novello}

\section{Connections between Weyl curvature, quantum potential and quantum entanglement}
Let us explore the relationships between Weyl curvature, quantum potential and quantum entanglement. For the 3D case,
the Weyl scalar curvature in (\ref{RiC2}) for the nD systems is reduced to
\begin{equation}\label{SR2}
\mathcal{R} =-2\lambda \left[\left(1-\frac{\lambda}{4}\right)
\frac{\nabla\Omega\cdot\nabla\Omega}{\Omega^2}
-\frac{\nabla^{2}\Omega}{\Omega}\right],
\end{equation}
where $\nabla=\mathbf{i}\frac{\partial}{\partial x_1}+\mathbf{j}\frac{\partial}{\partial x_2}+\mathbf{k}\frac{\partial}{\partial x_3}$ is the grade operator. Let $\lambda=4$, the Weyl scalar curvature is simplified further to
\begin{equation}\label{SR3}
\mathcal{R} =8 \frac{\nabla^{2}\Omega}{\Omega},
\end{equation}
which is consistent with the previous results.\cite{Novello,Licata1}
It implies that the Weyl flat $\mathcal{R} = 0$ corresponds to $\nabla^{2}\Omega=0$.
Note that $\quad Q = -\frac{\hbar^2}{2m}\frac{\nabla^{2}\Omega}{\Omega}$ and the quantum entropy defined by $S_Q=-\frac{1}{2}\ln \Omega$, thus,
$Q =(\nabla S_{Q})^{2}-\nabla^{2}S_{Q}$,\cite{Fiscaletti2}
the Weyl scalar curvature $\mathcal{R}$ can be also connected to quantum potential $Q$ and quantum entropy $S_Q$.
\begin{subequations}\label{RQ}
\begin{eqnarray}
   \mathcal{R} &=& -16\frac{m}{\hbar^2}Q,\\
   \mathcal{R} &=& 8\left[(\nabla S_{Q})^{2}-\nabla^{2}S_{Q}\right],\\
\end{eqnarray}
\end{subequations}
These results reveal the Weyl curvature can be interpreted to quantum probability, quantum potential and quantum entropy.
The Weyl scalar field plays a bridge to link geometry, quantum effect and quantum entropy together in the Q-wis geometry.
We can define a generalized force in Q-Wis spacetime by $F=-\nabla Q$, and then this force can be also expressed as
\begin{equation}\label{GF1}
F=\frac{\hbar^2}{16m}\nabla \mathcal{R}.
\end{equation}
It turns out that a force comes from the grade of Weyl scalar curvature in Q-Wis spacetime, which is equivalent to the force coming from the quantum potential.
This equivalence will be tested experimentally further.

Note that $\lambda=4$ for the 3D case and using the Weyl scalar curvature (\ref{SR3}), the Einstein-Weyl field equation (\ref{GE2}) in Q-Wis geometry
can be reduced to
\begin{equation}\label{GE2}
\frac{1}{\Omega}\frac{\partial^{2}\Omega}{\partial x_{i}\partial x_{j}}-\frac{3}{\Omega^2}\frac{\partial\Omega}{\partial x_{i}}\frac{\partial\Omega}{\partial x_{j}}
+g_{ij}\left(\frac{\nabla\Omega\cdot\nabla\Omega}{\Omega^2}+\frac{2m}{\hbar^2}Q\right) = 0.
\end{equation}
The Einstein-Weyl equation (\ref{GE2}) reveals the relationship between the Weyl scalar field and the quantum potential in Q-Wis geometry.

Moreover, the nonlinear Schr$\ddot{o}$dinger equation can be expressed by adding an intrinsic quantum potential. \cite{Wolfgang} In the Q-Wis geometry, the quantum potential can be equivalent to the Weyl scalar curvature. Thus, the nonlinear Schr$\ddot{o}$dinger equation can be rewritten as
\begin{equation}\label{NSCE}
i\hbar \frac{\partial \psi}{\partial t}=\left(-\frac{\hbar^2}{2m}\nabla^{2}+V+\frac{\hbar^2}{16m}\mathcal{R}\right)\psi.
\end{equation}
It implies that the Weyl scalar curvature can be interpreted not only the quantum potential, but also an intrinsic nonlinear effect in quantum mechanics.

For the 2D case, we find some differences from that in in the 3D case. Similarly
by substituting the quantum Weyl-Shannon entropy $f=-\lambda \ln \Omega$ to (\ref{RiC2}) and (\ref{SR2}), there is no a $\lambda$ solution that makes
a direct connection between the Weyl scalar curvature and quantum potential. The Weyl scalar curvature (\ref{SR1}) can be expressed to
\begin{subequations}\label{RQ2}
\begin{eqnarray}
\mathcal{R} & = & \lambda\left(\frac{\nabla^{2}\Omega}{\Omega}
-\frac{\nabla\Omega\cdot\bigtriangledown\Omega}{\Omega^2}\right),\\
\mathcal{R} &=& -\lambda\left(\frac{2m}{\hbar^2}Q+\frac{\nabla\Omega\cdot\bigtriangledown\Omega}{\Omega^2}\right), \\
\mathcal{R} &=& -\lambda\left(\nabla^{2}S_{Q}-\nabla S_{Q}\cdot \nabla S_{Q}+\frac{\nabla\Omega\cdot\bigtriangledown\Omega}{\Omega^2}\right),
\end{eqnarray}
\end{subequations}
where $\nabla=\mathbf{i}\frac{\partial}{\partial x_1}+\mathbf{j}\frac{\partial}{\partial x_2}$ is the grade operator for the 2D case.
We can also investigate the relationship between the Weyl scalar curvature and quantum potential in higher 3D cases. We find that there is also no $\lambda$ solution such that the Weyl scalar curvature directly connect to quantum potential for higher 3D cases.

Interestingly, these results imply that the relationships between the Weyl scalar curvature and quantum potential are dimension-dependent. The Weyl scalar curvature endowed with quantum potential works only in the 3D space.
This dimension-dependent correspondence between the Weyl scalar curvature and quantum potential inspires us a few of fundamental insights to nature:
\begin{itemize}
  \item \textit{It tells us why our space must be 3D if gravity must be quantized. In other words, quantum gravity must live in the 3D+1 spacetime.}
  \item \textit{It provides us a clue to quantize gravity by the Weyl geometry.}
  \item \textit{It could be applied in nanotechnology.}
\end{itemize}

\begin{figure}[pb]
\begin{center}
\includegraphics[scale=0.5]{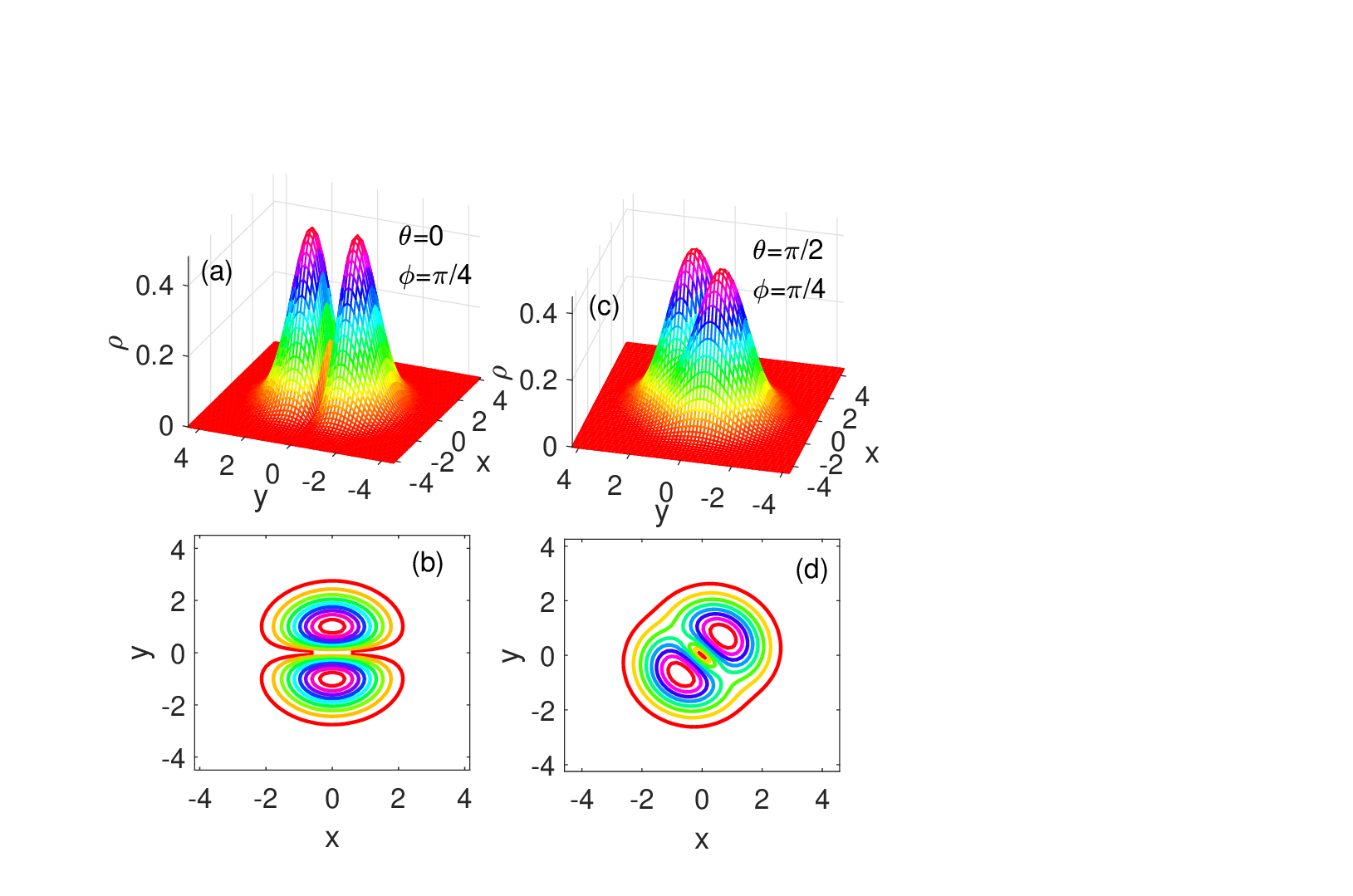}
\caption{Online color: The profiles of the probability densities in the separate and entangled states. (a) The profile of the probability density in the separate state $\theta=0$ and (b) its contour map. (c) The profile of the probability density in the maximum entangled state $\theta=\pi/2$ and (d) its contour map.}
	\label{Fig1}
\end{center}
\end{figure}

\section{A typical example: two orthogonal linear oscillators}
To explore the relationship between the geometrical properties and quantum entanglement of quantum systems based on the formulation of the Q-Wis geometry, we study two 1D mutually orthogonal linear oscillators to form a 2D system as a typical example, in which there is no interaction. The Hamiltonian can be written as
\begin{equation}\label{H2}
H=H_{1}+ H_{2}=\frac{\widehat{p}^{2}_{x}}{2m}+\frac{m\omega^2}{2}x^2
+\frac{\widehat{p}^{2}_{y}}{2m}+\frac{m\omega^2}{2}y^{2}.
\end{equation}
We investigate the systems in the the Bell-entangled state between the ground state in the oscillator $1$ and the first excited state in the oscillator $2$.
The wave function of the  Bell entangled state is expressed as
\begin{equation}\label{WF1}
\Psi(x,y)=\cos\frac{\theta}{2}\psi^{(1)}_{0}(x)\psi^{(2)}_{1}(y)+e^{i\varphi}\sin\frac{\theta}{2}\psi^{(1)}_{1}(x)\psi^{(2)}_{0}(y),
\end{equation}
where $\psi^{(i)}_{n}$ represents the wave function of the oscillator $i$ in the $n-$ state, which are
\begin{subequations}\label{WF2}
\begin{eqnarray}
\psi_{0}(x) &=& N_{0}e^{-\alpha^{2}x^2/2}, \\
\psi_{1}(y) &=& N_{1}2\alpha x e^{-\alpha^{2}y^2/2},
\end{eqnarray}
\end{subequations}
where $\alpha=\sqrt{\frac{m\omega}{\hbar}}$ and $N_{n}=\sqrt{\frac{\alpha}{\sqrt{\pi}2^{n}n!}}$ is the normalized constants for the n-state. The density of the particle probability is $\rho(x,y)=\Psi^{*}(x,y)\Psi(x,y)$.
The bipartite quantum entanglement can be measured by the Neumann quantum entropy of the reduced density matrix,
\begin{equation}\label{NE1}
S_{N}=-\textrm{Tr}(\widehat{\rho}_{r}\ln \widehat{\rho}_{r}),
\end{equation}
where $\widehat{\rho}_{r}=\textrm{Tr}_{1(2)}(\widehat{\rho})$ is the reduced density matrix and $\widehat{\rho}=|\Psi\rangle\langle\Psi|$ is the density matrix of the bipartite system. The quantum entanglement depends on $\theta$, but is independent of $\varphi$. Thus, we set $\varphi=\pi/4$ in the following numerical investigations and $\theta=0, \pi/2$ for the typical separate and maximum entangled states, respectively.

Firstly let us look at the profile of the probability density in Fig.1. We compare two typical cases, namely $\theta=0$ is the separate state and $\theta=\pi/2$ is the maximum entangled state. It can be seen that the probability density in the separate state is separate and it is entangled mixing in the maximum entangled state.

\begin{figure}[pb]
\begin{center}
\includegraphics[scale=0.5]{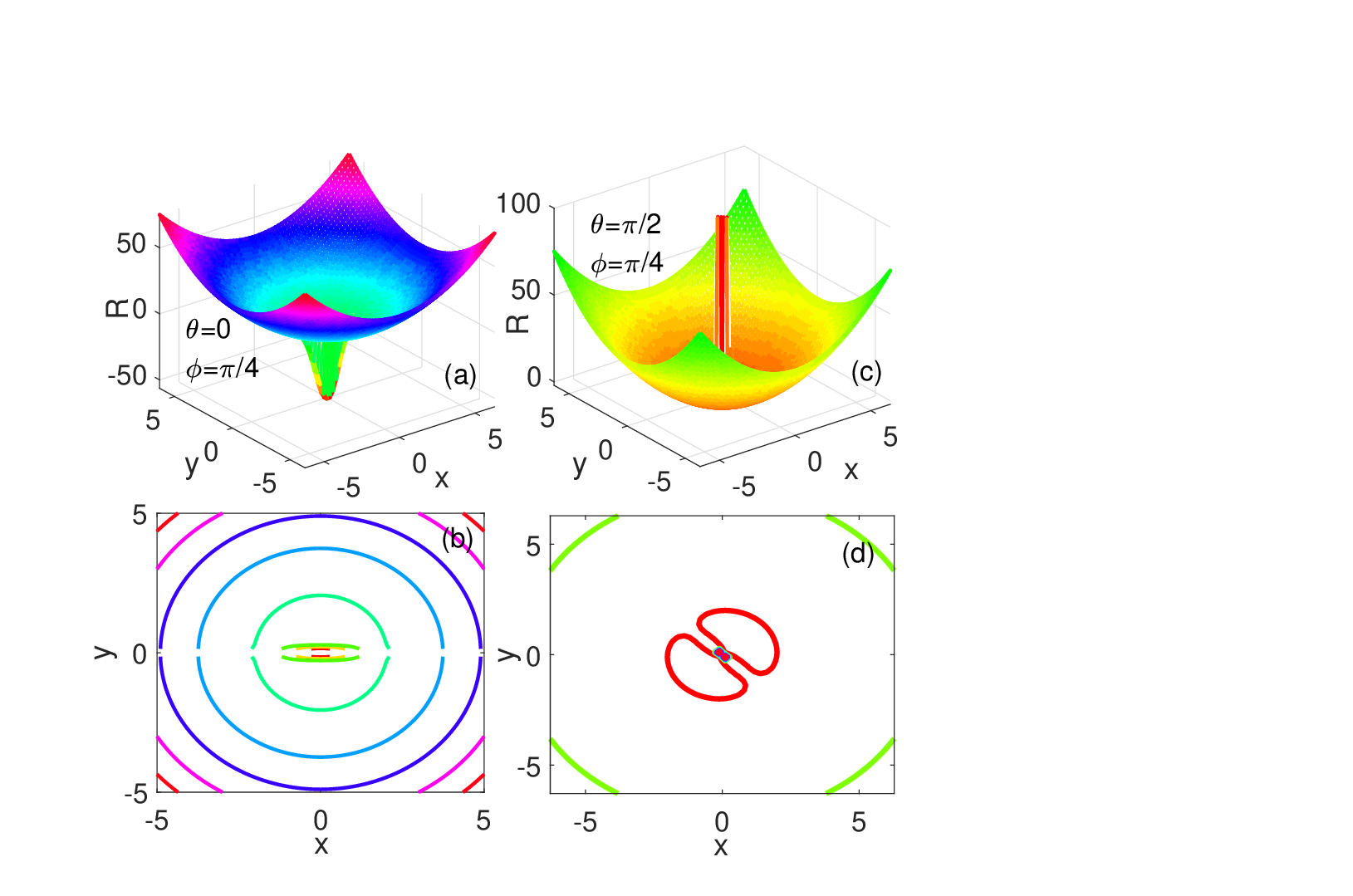}
\caption{Online color: The comparison of the Weyl scalar curvatures between the separate $\theta=0$ and the maximum entangled state $\theta=\pi/2$. (a) The profile of the Weyl scalar curvature and (b) its contour map. (c) The profile of the Weyl scalar curvature and (d) its contour map.}
	\label{Fig2}
\end{center}
\end{figure}

\subsection{Weyl scalar curvature versus Entanglement}
As a typical geometrical variables, we investigate the Weyl scalar curvature in the typical cases of $\theta=0$ and $\pi/2$, which correspond to the separate state and the maximum entanglement of the bipartite system of two orthogonal linear oscillators.

Figure 2 shows the profile of the Weyl scalar curvature in the separate state $(\theta=0)$ of two orthogonal linear oscillators. We can see from Fig.2(a) that the Weyl curvature is negative near the original point $(0,0)$ and shows a pit along the $x$ axis. The corresponding contour map is in Fig.2 (b). For the maximum entangled state, $\theta=\pi/2$, the Weyl curvature becomes positive and shows a peak near the original point (see Fig.2(c) and the pattern is symmetric to the $\pi/4$ direction. This change of the Weyl scalar curvature for the separate and entangled states provides a geometrical signal to capture the quantum entanglement.

\begin{figure}[pb]
\begin{center}
\includegraphics[scale=0.5]{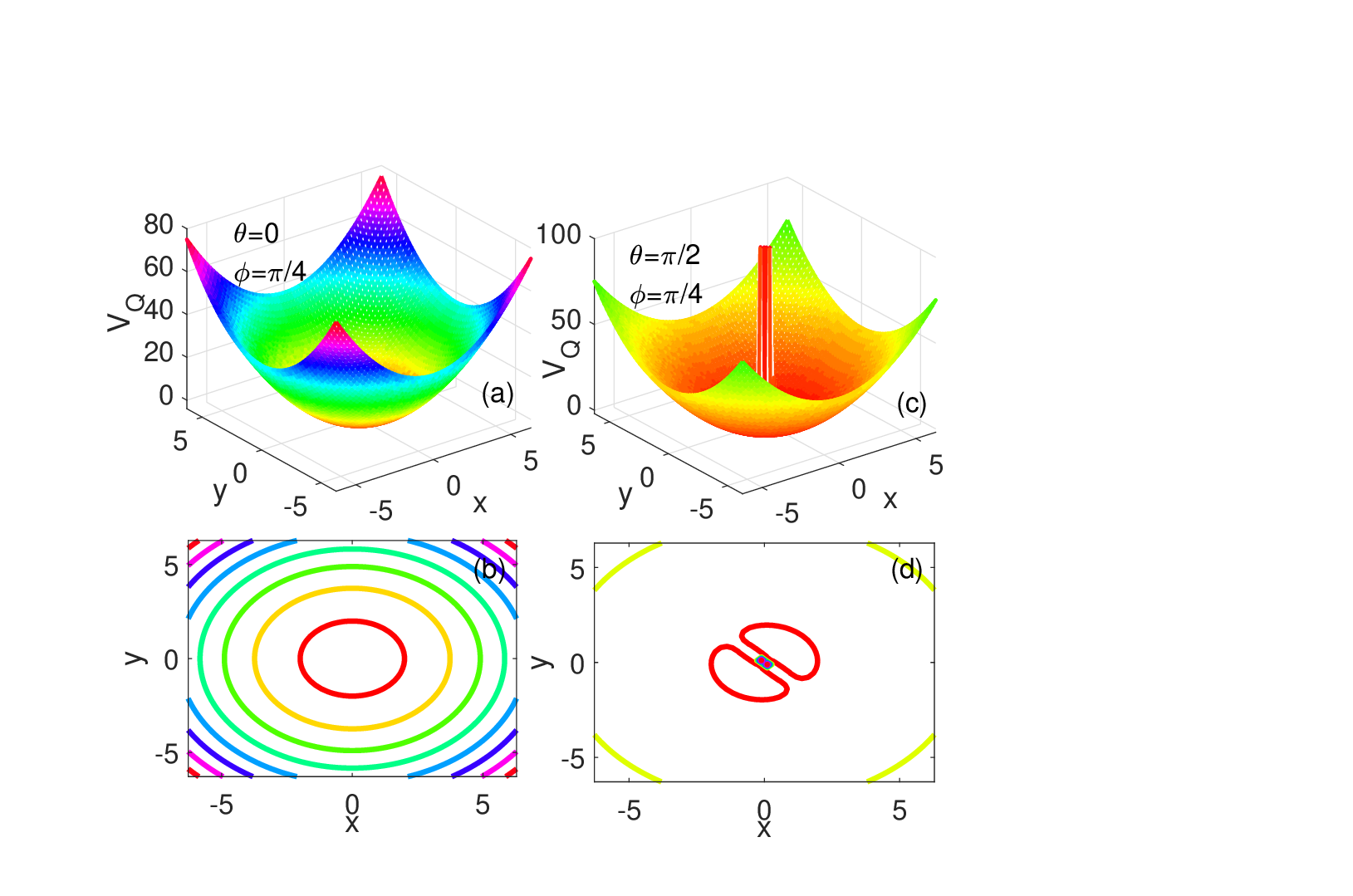}
\caption{Online color: The comparison of the quantum potentials between the separate and entangled states. (a) The quantum potential in the separate state and (b) its contour map. (c) The quantum potential in the maximum entangled state and (d) its contour map.}
	\label{Fig3}
\end{center}
\end{figure}

\subsection{Quantum potential versus Entanglement}
Similarly, we plot the profiles of the quantum potentials in the separate and entangled states in Fig.3 to compare the behaviors of the quantum potentials in the separate and entangled states. It can been seen from the Figs.3(a) that the quantum potential in the separate state is a smooth paraboloid. In the maximum entangled state in Figs.3 (c) the quantum potential near the original point shows a sharp peak like the Weyl scalar curvature. This indicates that the quantum potential depends on the quantum entanglement.

By comparing the behaviors of the Weyl scalar curvature and quantum potential in the separate state and the maximum entangled state, we can find their differences near the original points. The Weyl scalar curvature becomes negative near the original point, which comes from the second term in (\ref{RQ2}b).

\begin{figure}[pb]
\begin{center}
\includegraphics[scale=0.4]{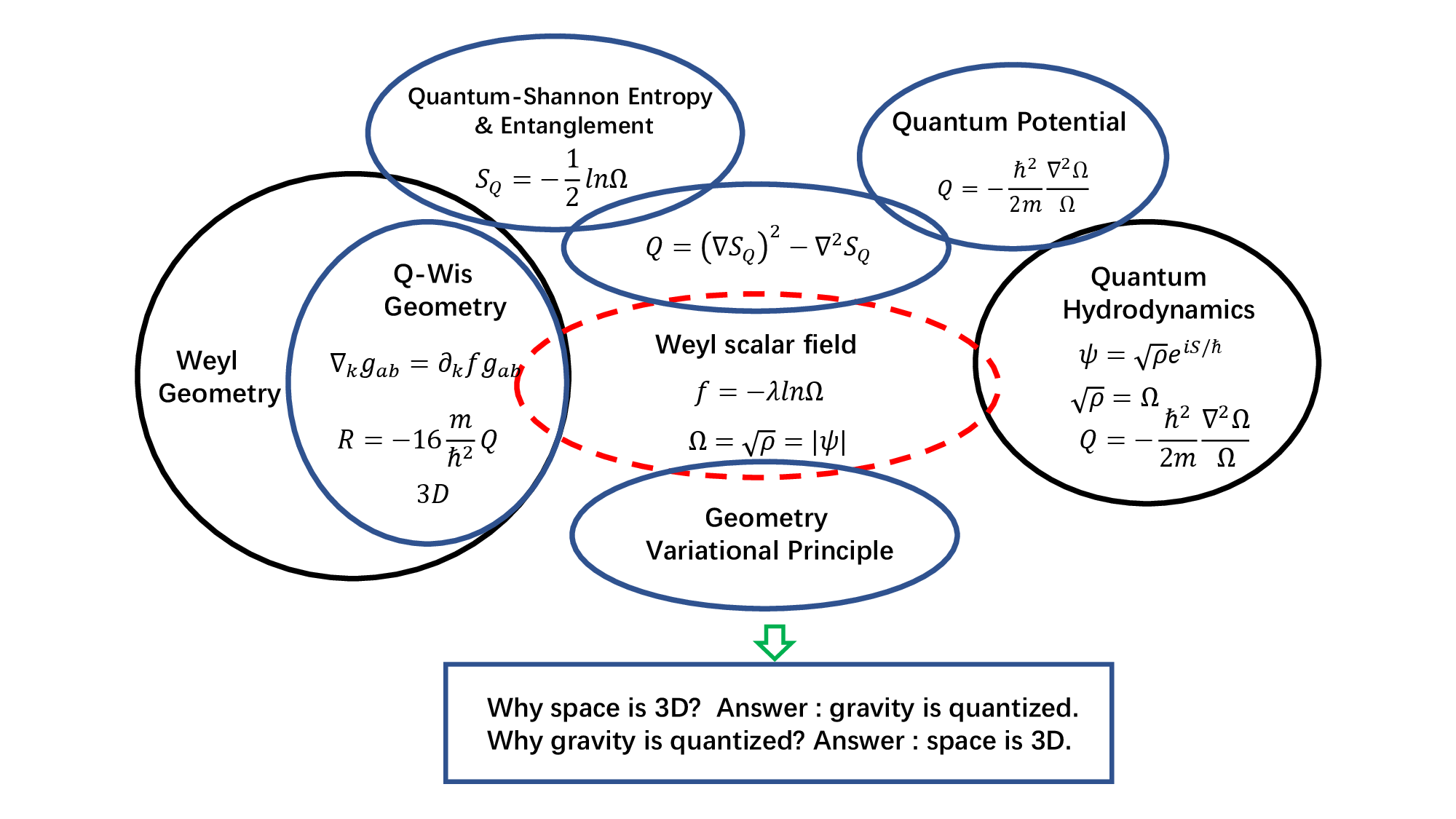}
\caption{Online color: The summary of the connections between Weyl geometry and quantum mechanics.}
	\label{Fig4}
\end{center}
\end{figure}

\section{Conclusions and outlooks}
The geometric description of quantum systems provides a novel view to quantum potential and quantum entanglement.
The Weyl scalar field induced by the Weyl geometry play a bridge role connecting the Weyl scalar curvature and quantum potential as well as quantum entanglement. We have generalized the formulation of the Q-Wis geometry to the nD case and found that the connection between the Weyl scalar curvature and the quantum potential is dimension-dependent.

In the 3D case, the Weyl scalar curvature can be equivalent to quantum potential by identifying $\lambda=4$.
This connection gives us a novel insight to Weyl geometry and quantum effects.
The generalized Einstein-Weyl field equation (\ref{GE2}) and the generalized Schr$\ddot{o}$dinger equation (\ref{NSCE}) in 3D vacuum case
reveals the relationship between Weyl geometry and quantum potential. These findings opens a novel view to the correspondence between Weyl scalar curvature and quantum potential as well as quantum entanglement, which inspires deeper relationships between geometry and quantum effect of spacetme.

It should be emphasized that beyond the 3D there is no a $\lambda$ solution such that the Weyl scalar curvature is directly equivalent to quantum potential. This dimension-dependent relationship between the Weyl geometry and quantum mechanics reveals some clues to quantize gravity.

In the journey to unify gravity and quantum mechanics, one developed different approaches
such as canonical quantum gravity,\cite{THOMAS,CLAUS} quantum geometry\cite{Eduard,Matthew} and string theory,\cite{JOSEPH} to set up a compatible framework for understanding
gravity, spacetime and quantum mechanics. However, a lot of puzzles still troubles physicist,\cite{Lee} such as how emergence of spacetime and why higher dimensional space (9- or 10-dimensional space for string theory) hide behind our observations.\cite{Ralph}  Why do the theoretical frameworks have to be generalized to higher spatial space for unification of gravity and quantum mechanics but our observations must be constrained in the 3D space?.
Our findings inspire us to propose a hypothesis on the relationship between spacetime, gravity and quantum mechanics:

{\it Spacetime lives in Weyl geometry and Weyl scalar curvature is endowed with quantum potential, which
answer a natural puzzle why our space is three and gravity must be quantized and quantum gravity must live the 3D+1 spacetime.}

This hypothesis tell us not only why the spatial dimension is three but also a clues on how to compatibly unify gravity and quantum mechanics based on the Weyl geometry.

By numerical investigating two 1D orthogonal linear oscillators, we have revealed the relationships between the Weyl scalar curvature, quantum potential, quantum entanglement, and quantum potential. We have found that the Weyl scalar curvature shows quite different behaviors in the separate state and the maximum entangled state. The Weyl scalar curvature has a negative pit in the separate state, but shows a positive peak in the maximum entangled state near the original point. This can be regarded as a geometric signal to detect quantum entanglement.

These results imply some connections between the Weyl geometry and quantum entanglement. The geometry structure could emerge from quantum effect. Inversely the geometrical characteristics could induces quantum uncertainty.
The Weyl scalar curvature allows us define the Weyl length by $L_{W}=\frac{1}{\sqrt{\mathcal{R}}}$.\cite{Fiscaletti1,Licata3}
The Weyl length depends on the quantum potential or quantum entanglement, which
implies the momentum quantum fluctuation based on the uncertainty principle, $\triangle p \sim \hbar/L_{W}$ coming from the Weyl geometry.

Fig. 4 summarizes the connections between Weyl geometry, quantum potential, quantum entropy and quantum entanglement.
The Weyl-Shannon quantum entropy as an analog with the Weyl scalar field can be connected to Fisher information and gravity.\cite{Carroll,Fiscaletti2,Licata3,Castro} This inspires a more general equivalent principle, geometry could be equivalent to quantum randomness in some sense, in which the Weyl scalar field plays a bridge role to connect the curvature and quantum potential or quantum entanglement.
This connection could induce a deeper unification of quantum randomness, quantum entanglement, gravity and information, which also answers two natural puzzles why our space is 3D and why gravity should be quantized.





\end{document}